\documentclass[a4paper]{jpconf}

\usepackage{graphicx}
\usepackage{amsmath,amssymb}
\def\ga{{\mbox {${\alpha}$}}}

\def\ge{{\mbox {${\epsilon}$}}}
\def\gm{{\mbox {${\gamma}$}}}
\def\gth{{\mbox {${\theta}$}}}

\def\lam{{\mbox {${\lambda}$}}}
\def\Gm{{\mbox {${\Gamma}$}}}
\def\Lam{{\mbox {${\Lambda}$}}}
\def\Sig{{\mbox {${\Sigma}$}}}
\def\sg{{\mbox {${\sigma}$}}}
\def\pr{\prime}
\def\pd{\partial}
\def\half{\frac{1}{2}}

\def\rr{\rangle\!\rangle}
\def\ll{\langle\!\langle}

\begin{document}

\title{Algebraic description of external and internal attributes of fundamental fermions}

\author{Ikuo S. Sogami}

\address{Maskawa Institute for Science and Culture, Kyoto Sangyo University, Kita-Ku,
Kyoto 603-8555, JAPAN}

\ead{sogami@cc.kyoto-su.ac.jp}

\begin{abstract}
To describe external and internal attributes of fundamental fermions, a theory of
multi-spinor fields is developed on an algebra, a {\it triplet algebra}, which consists
of all the triple-direct-products of Dirac $\gamma$-matrices. The triplet algebra is
decomposed into the product of two subalgebras, an external algebra and an internal algebra,
which are exclusively related with external and internal characteristic of the multi-spinor
field named {\it triplet fields}. All elements of the external algebra which is isomorphic
to the original Dirac algebra $A_\gm$ are invariant under the action of permutation group
S$_3$ which works to exchange the order of the $A_\gm$ elements in the triple-direct-product.
The internal algebra is decomposed into the product of two $4^2$ dimensional algebras,
called the family and color algebras, which describe the family and color degrees of freedom.
The family and color algebras have fine substructures with {\lq}{\lq}trio plus solo{\rq}{\rq}
(\,3\,+\,1\,) conformations which are irreducible under the action of S$_3$. 
The triplet field has trio plus solo family modes with ordinary tricolor quark and
colorless solo lepton components. To incorporate the Weinberg-Salam mechanism,
it is required to introduce two types of triplet fields, a left-handed doublet
and right-handed singlets of electroweak iso-spin. It is possible to qualify
the Yukawa interaction and to make a new interpretation of its coupling constants naturally
in an intrinsic mechanism of the triplet field formalism. The ordinary Higgs mechanism
leads to a new type of the Dirac mass matrices which can explain all data of quark sector
within experimental accuracy.
\end{abstract}

\section{Introduction}
Fundamental fermions, quarks and leptons, exist in repetitional chiral modes with color
and electroweak symmetries. To describe this rich structure as a whole, it is necessary
to prepare a  flexible theoretical framework with large capacity. In the present article,
a theory of multi-spinor fields is developed on an algebra, a {\it triplet algebra}, which 
consists of all the triple-direct-products (TDPs) of Dirac $\gamma$-matrices. The triplet
algebra is decomposed into the product of two subalgebras, an external algebra and an
internal algebra, which are exclusively related with external and internal characteristic
of fundamental fermions. The multi-spinor field, which is called a {\it triplet field},
forms reducible representation of the Lorentz group including spin $\frac{1}{2}$ component
fields with a variety of internal attribute and symmetry.

All elements of the external algebra which is isomorphic to the original Dirac algebra
$A_\gm$ are invariant under the action of permutation group S$_3$ which works to exchange
the order of the $A_\gm$ elements in the TDP. The internal algebra is
decomposed into the product of two $4^2$ dimensional subalgebras, called the family and
color algebras, which describe the family and color degrees of freedom, respectively. Correspondingly, the triplet field possesses independent component fields of maximally
quartet family modes and quartet color modes.

We impose the conditions that the family and color algebras have fine substructures which
are irreducible under the action of the permutation group S$_3$. This condition works to
degrade the family and color algebras which are responsible for the quartet structures
into fine algebras with {\lq}{\lq}trio plus solo{\rq}{\rq} (\,3\,+\,1\,) conformations.
As a result, we are able to interpret that the component fields of the triplet field
are composed of the ordinary three families and an additional fourth family of the tricolor
quark and colorless lepton modes.   

To reproduce all the successful results of the standard model and go beyond it, we
introduce two types of triplet fields, a left-handed doublet and right-handed singlets
of electroweak iso-spin, and apply the Weinberg-Salam theory originally formulated for
the electron and its associated neutrino to these triplet fields. The ordinary
gauge fields and Higgs field of electroweak SU(2) symmetry are assumed to interact
with these triplet fields, and the gauge interactions of color SU(3) symmetry
({\lq}{\lq}baryon minus lepton{\rq}{\rq} $B-L$ charge) are set to take effects
between internal quark modes (for quark and lepton modes) of the triplet fields.
Thanks to the {\lq}{\lq}trio plus solo{\rq}{\rq} scheme for the family structure,
the Yukawa interactions are block diagonalized so that ordinary trio quarks and
trio leptons interact among themselves. The solo fermions in the fourth family exist
independently of the ordinary trio fermions without observable mixing effects.

The Lagrangian density of the fermion sector is constructed with the triplet fields.
By substituting the mode decompositions of the triplet fields, we express the density
in terms of component fields of quarks and leptons. One superior aspect of this formalism
is calculability of the Yukawa coupling constants as matrix elements between different
family modes. It is possible to qualify the Yukawa interaction and to make a new
interpretation of its coupling constants with CP-violating phase naturally in an intrinsic
mechanism of the triplet field formalism. The ordinary Higgs mechanism results in
new and unique Dirac mass matrices which can explain all of the experimental data
of the quark mass spectra and weak mixing matrix.

We construct explicitly the triplet algebra and its subalgebra for the external spacetime
symmetry and internal attributes of fundamental fermions in \S 2. Characteristic of
the subalgebra for internal properties of fundamental fermions is closely examined
in \S 3 and \S 4. First, chirality operators acting on the triplet field are decomposed
into the sums of the projection operators which can be interpreted to discriminate
repetitional degrees of freedom for family structure. Finding the elements which can
describe the mixings between the interaction and mass-eigenstate modes, we construct
the family algebra with the trio plus solo conformation. Second, the algebra for
color symmetry is determined in the set of the elements which are commutative both
with the chirality operators and all elements of the algebra for external symmetry. 
We introduce the triplet field forming a multiple representation of the Lorentz group
in \S 5. In the present scheme, the triplet field has no degrees of freedom for
the symmetry of electroweak interaction. In \S 6, we reformulate the standard model
by introducing the set of chiral triplet fields of the left-handed doublet and
the right-handed singlets of the Weinberg-Salam symmetry. Explicit form of the Yukawa
interaction is given and the Dirac mass matrix with new unique structure is derived.
Possible implications of the triplet field formalism for and beyond the standard model
are discussed in \S 7.

\section{Triplet algebra and subalgebras for external and internal attributes}
The Dirac's $\gm$-matrices $\gm_\mu\ (\mu = 0,\ 1,\ 2,\ 3)$ satisfying the Clifford anticommutation relations 
$\gm_\mu\gm_\nu + \gm_\nu\gm_\mu = 2\eta_{\mu\nu}1$ with
$(\, \eta_{\mu\nu}\, ) = {\rm diagonal}\, ( 1, -1, -1, -1 )$ are introduced solely as the mathematical quantities which carry no direct physical meaning. Hermite conjugate of $\gm_\mu$ is defined by $\gm_\mu^\dagger = \gm_0\gm_\mu\gm_0$. The $\gm$-matrices generate the 16 dimensional algebra, named a {\it Dirac algebra}, as
\begin{equation}
 A_{\gm} =\ \langle \gm_\mu \rangle =\ 
 \{1,\, \gm_\mu,\, \sigma_{\mu\nu},\, \gm_5\gm_\mu,\, \gm_5 \}
 \label{DiracAlg}
\end{equation}
where $\sigma_{\mu\nu} =\frac{i}{2}(\gm_\mu\gm_\nu-\gm_\nu\gm_\mu)$ and
$\gm_5 = i\gm_0\gm_1\gm_2\gm_3 =  \gm^5$.

Let us call the TDP of the bases
$1,\, \gm_\mu,\, \sigma_{\mu\nu},\, \gm_5\gm_\mu$ and $\gm_5$ of $A_{\gm}$ a {\it primitive triplet}
and the TDP of arbitrary elements of $A_{\gm}$ a {\it triplet}. Then,
the {\it triplet algebra} $A_T$ is defined as the $16^3$ dimensional algebra spanned by all
the linear combinations of triplets. In other words, the triplet algebra $A_T$ is generated
in terms of the 12 primitive triplets $\gm_\mu\otimes 1\otimes 1,\ 1\otimes\gm_\mu\otimes 1$
and $1\otimes 1\otimes\gm_\mu$ as follows:
\begin{equation}
   A_T = \langle \gm_\mu\otimes 1\otimes 1,\ 
   1\otimes\gm_\mu\otimes 1,\ 1\otimes 1\otimes\gm_\mu\rangle.
\end{equation}
The transpose (Hermite conjugate) of the primitive triplet is defined by the
TDP of its transposed (Hermite conjugate) components of $A_\gm$ and the
trace of the triplet is set to be the product of the traces of its $A_{\gm}$ components. 

The triplet algebra $A_T$ is large enough to form a variety of realizations of the Lorentz
group representation. To obtain such kind of multiple spinor representation that accomodates
solely the spin $\frac{1}{2}$ states, we note that the primitive triplets
\begin{equation}
   \Gm_\mu = \gm_\mu\otimes\gm_\mu\otimes\gm_\mu\qquad
   (\mu = 0,\,1,\,2,\,3)
\end{equation}
satisfy the Clifford relations~\cite{Sogami1,Sogami2}
\begin{equation}
   \Gm_\mu\Gm_\nu + \Gm_\nu\Gm_\mu = 2\eta_{\mu\nu}I
\end{equation}
where $I = 1\otimes 1\otimes 1$. Hermite conjugate of $\Gm_\mu$ is defined by
$\Gm_\mu^\dagger = \Gm_0\Gm_\mu\Gm_0$ and the trace is calculated as
${\rm Tr}\,\Gm_\mu = 0$ and ${\rm Tr}\left( \Gm_\mu\Gm_\nu \right) = 64 \eta_{\mu\nu}$.
With these new elements, we are now able to construct a subalgebra for external attributes
of the triplet field, the {\it external algebra} $A_{ex}$, as follows:
\begin{equation}
A_{ex} =\  \langle \Gm_\mu \rangle
=\ \{1,\, \Gm_\mu,\, \Sigma_{\mu\nu},\, \Gm_5\Gm_\mu,\, \Gm_5 \}
\end{equation}
where 
\begin{equation}
  \Sigma_{\mu\nu}= -\frac{i}{2}(\Gm_\mu\Gm_\nu - \Gm_\nu\Gm_\mu)
 = \sigma_{\mu\nu}\otimes\sigma_{\mu\nu}\otimes\sigma_{\mu\nu}
\end{equation}
and
\begin{equation}
  \Gm_5 = -i\Gm_0\Gm_1\Gm_2\Gm_3 = \Gm^5
        = \gm_5\otimes\gm_5\otimes\gm_5.
\end{equation}
Evidently, $A_{ex}$ is isomorphic to the original Dirac algebra $A_\gm$.

It is readily proved that the operators
\begin{equation}
   M_{\mu\nu} = \frac{1}{2}\Sigma_{\mu\nu}
   \label{Lorentzgenerator}
\end{equation}
are subject to the commutation relations of the Lie algebra for the orthogonal group O(1,\,3)
and $M_{\mu\nu}$ and $\Gm_\lam$ satisfy the relations
$[ M_{\mu\nu},\ \Gm_\lambda ] = i\eta_{\lambda\nu} \Gm_{\mu}-i\eta_{\lambda\mu} \Gm_{\nu}$.
Accordingly, it is possible to postulate that the operators $M_{\mu\nu}$ generate the Lorentz transformations in our 4 dimensional spacetime and that the subscripts of operators consisting
of $\Gm_\mu$ refer directly to the superscripts of the spacetime coordinates $\{\,x^\mu\,\}$
of the 4 dimensional world where we exist as observers. Therefore, it is naturally defined
that the raising and lowering operations of index of $\Gm_\mu$ by the metric $\eta_{\mu\nu}$ as
$\Gm^\mu = \eta^{\mu\nu}\Gm_\nu$ and $\Gm_\mu = \eta_{\mu\nu}\Gm^\nu$.

The internal attributes of fundamental fermions should be fixed independently of the inertial
frame of reference in which observations are made. This means that the generators specifying
their internal attributes must be commutative with the elements of the subalgebra,
$\{\Sig_{\mu\nu}\}$, for the Lorentz transformations. Accordingly, we are led naturally to
postulate that the {\it internal algebra}, $A_{in}$, is defined by the {\it centralizer}
of $\{\Sig_{\mu\nu}\}$ of the algebra for the Lorentz transformation as follows:
\begin{equation}
   A_{in} = C(\{\Sig_{\mu\nu}\}: A_T) = \{ X \in A_T : [\, X,\ \Sigma_{\mu\nu} \,] = 0 \}.
\end{equation}
The primitive triplet of this centralizer is proved to consist of either an even number
or an odd number of component $\gm_\mu$-matrix for all $\mu$~\cite{Sogami1}.
Consequently, the internal algebra is determined to be
\begin{equation}
   A_{in} = \langle\ 1\otimes \gm_\mu\otimes \gm_\mu,\  
   \gm_\mu\otimes 1\otimes\gm_\mu,\ \gm_5\otimes\gm_5\otimes\gm_5\ \rangle .
   \label{Ain}
\end{equation}
Now, the triplet algebra is shown to have the decomposition
\begin{equation}
   A_T = A_{ex} A_{in},\quad  A_{ex}\cap A_{in} = \{ \Gm_5\}.
      \label{whereisGm5}
\end{equation}
The element $\Gm_5$ belongs not only to the external algebra but also to the internal
algebra. This is a crucial aspect of the present theory which makes it possible to realize
the repetitional family structure as an internal attribute of the triplet field.

As is evident from the construction so far, the component $\gm$-matrices carry no direct physical
meaning and regarded as playing only the role of {\it alphabets}. It is selected elements
of the triplet algebra $A_T$ that play the role of {\it codons} carrying the direct physical
meanings.

\section{Subdivision of chirality operators and repetitional modes of family structure}
Our next task is to examine detailed structures of the internal algebra $A_{in}$
in (\ref{Ain}) and to inquire what sorts of internal attributes of the
fundamental fermions can be inscribed on the algebra. For its purpose, we should notice
the point that complexties in flavor physics rise from mixture of freedom of families and
chiralities.   

It is straightforward to verify that the chirality operators
\begin{equation}
   L = \frac{1}{2}(I-\Gm_5), \ \ R = \frac{1}{2}(I+\Gm_5)
\end{equation}
have the following subdivisions in the internal algebra $A_{in}$ as
\begin{equation}
  L = \ell\otimes r\otimes r + r\otimes\ell\otimes r
    + r\otimes r\otimes\ell + \ell\otimes\ell\otimes\ell
  \label{Loperator}
\end{equation}
and
\begin{equation}
  R = r\otimes \ell\otimes\ell + \ell\otimes r\otimes\ell
    + \ell\otimes\ell\otimes r + r\otimes r\otimes r 
  \label{Roperator}
\end{equation}
where
\begin{equation}
   \ell = \frac{1}{2}(1-\gm_5),\ \ r = \frac{1}{2}(1+\gm_5).
\end{equation}
Note that this subdivision cannot be realized in the algebra $A_{ex}$.

To give independent physical roles and meanings to the subdivided terms in (\ref{Loperator})
and (\ref{Roperator}), let us introduce the following operators in $A_{in}$ as
\begin{equation}
 \left\{\  
  \begin{array}{ll}
  \Pi_{1L} = \ell\otimes r\otimes r,\ &
  \Pi_{1R} = r\otimes \ell\otimes\ell, \\
 \noalign{\vskip 0.2cm}
  \Pi_{2L} = r\otimes\ell\otimes r,\ &
  \Pi_{2R} = \ell\otimes r\otimes\ell, \\
 \noalign{\vskip 0.2cm}
  \Pi_{3L} = r\otimes r\otimes\ell,\ &
  \Pi_{3R} = \ell\otimes\ell\otimes r,  \\
 \noalign{\vskip 0.2cm}
  \Pi_{4L} = \ell\otimes\ell\otimes\ell,\ &
  \Pi_{4R} = r\otimes r\otimes r, \\
  \end{array}
 \right.
\end{equation}
which are chiral projection operators obeying the relations
\begin{equation}
   \Pi_{ih}\Pi_{jh^\prime} = \delta_{ij}\delta_{hh^\prime}\Pi_{ih},\ \ \sum_{ih}\Pi_{ih}=I 
\end{equation}
for $i,\,j = 1,\,2,\,3,\,4$ and $h,\,h^\pr = L,\,R$. Next, we define the non-chiral projection operators
\begin{equation}
     \Pi_i = \Pi_{iL} + \Pi_{iR}
\end{equation}
satisfying the relations
\begin{equation}
   \Pi_i\Pi_j = \delta_{ij}\Pi_i, \ \ \sum_{i}\Pi_{i}=I.
\end{equation}
Note that the chiral projection operators can be reproduced from the non-chiral projection
operators as
\begin{equation}
   \Pi_{iL}=L\Pi_i,\quad \Pi_{iR}=R\Pi_i.
   \label{nonchiraltochiral}
\end{equation}

Here we hypothesize that this decomposition of the chirality operators explains the origin of
the family structure of fundamental fermions. There are maximally four family degrees of
freedom in this scheme. However, experimental data have established now that the family number
of ordinary quarks and leptons are three in low energy regime of flavor physics. To incorporate
this observed characteristics, let us notice the different behaviors of the projection operators
$\Pi_i$ under actions of the discrete group S$_3$ which works to exchange the order of the
elements of $A_\gm$ in the triple-direct product. Under the action of this group, the three
operators $\Pi_i\ (i=1, 2, 3)$ are changed with each other and the operator $\Pi_4$ remains
invariant. It is natural and relevant to make use of this feature to divide the quartet into
{\lq}{\lq}trio and solo{\rq}{\rq} sectors for the family structure.

The projection operators to the trio and solo sectors are constructed in the forms
\begin{equation}
  \Pi_{(t)} = \Pi_1 + \Pi_2 + \Pi_3
        = \frac{1}{4}\left(3I - 1\otimes\gm_5\otimes\gm_5 
                             - \gm_5\otimes 1\otimes\gm_5
                             - \gm_5\otimes\gm_5\otimes 1 \right)
        \label{triof}
\end{equation}
and
\begin{equation}                     
  \Pi_{(s)} = \Pi_4
      = \frac{1}{4}\left(I + 1\otimes\gm_5\otimes\gm_5 
                                + \gm_5\otimes 1\otimes\gm_5
                                + \gm_5\otimes\gm_5\otimes 1 \right)
         \label{solof}
\end{equation}
which are subject to the relations $\Pi_{(a)}\Pi_{(b)} = \delta_{ab}\Pi_{(a)}$ for
$a, b = t, s$. We postulate here that the ordinary quarks and leptons belong to the
trio sector and a fourth family of additional quark and lepton forms the solo sector. 

Quarks and leptons can exist in either modes of the interaction and the mass eigenstates.
To allow transmutations among two modes, it is necessary to secure a sufficient number of
ingredients in the algebra for the trio sector, {\it trio algebra}. For its purpose,
let us introduce the three elements of $A_{\gm}$ as
\begin{equation}
   \sigma_1 = \gm_0,\ \sigma_2 = i\gm_0\gm_5, \ \sigma_3 = \gm_5
   \label{sigma}
\end{equation}
which are subject to the multiplication rules of the Pauli algebra, {\it i.e.},
$\sigma_a\sigma_b = \delta_{ab}I + i\ge_{abc}\sigma_c$. Taking the TDPs
of these elements, we are able to construct the bases for the trio algebra as follows:
\begin{equation}
\left\{\ 
  \begin{array}{l}
    \pi_1 = \half\left(\sigma_1\otimes\sigma_1\otimes 1
          + \sigma_2\otimes\sigma_2\otimes 1\right), \\
             \noalign{\vskip 0.3cm}
    \pi_2 = \half\left(\sigma_1\otimes\sigma_2\otimes\sigma_3
          - \sigma_2\otimes\sigma_1\otimes\sigma_3\right),\\
          \noalign{\vskip 0.3cm}
    \pi_3 = \half\left(1\otimes\sigma_3\otimes\sigma_3
          - \sigma_3\otimes 1\otimes\sigma_3\right),  \\
          \noalign{\vskip 0.3cm}
    \pi_4 = \half\left(\sigma_1\otimes 1\otimes\sigma_1
          + \sigma_2\otimes 1\otimes \sigma_2\right), \\ 
             \noalign{\vskip 0.3cm}
    \pi_5 = \half\left(\sigma_1\otimes\sigma_3\otimes\sigma_2
          - \sigma_2\otimes\sigma_3\otimes\sigma_1\right),\\
          \noalign{\vskip 0.3cm}
    \pi_6 = \half\left(1\otimes\sigma_1\otimes\sigma_1
          + 1\otimes\sigma_2\otimes\sigma_2\right), \\ 
             \noalign{\vskip 0.3cm}
    \pi_7 = \half\left(\sigma_3\otimes\sigma_1\otimes\sigma_2
          - \sigma_3\otimes\sigma_2\otimes\sigma_1\right),\\
          \noalign{\vskip 0.3cm}
    \pi_8 = \frac{1}{2\sqrt{3}}\left(1\otimes\sigma_3\otimes\sigma_3
          + \sigma_3\otimes 1\otimes\sigma_3 
          -2\sigma_3\otimes\sigma_3\otimes 1\right) .
\end{array}
\right.
 \label{pi}
\end{equation}
It is somehow laborious but straightforward to verify that these eight operators belonging
to $A_{in}$ satisfy the commutation relations
\begin{equation}
        [\pi_j,\ \pi_k] = 2f_{jkl}\pi_l
        \label{comsu3pi}
\end{equation}
and the anticommutation relations
\begin{equation}
       \{\pi_j,\ \pi_k\} = \frac{4}{3}\delta_{jk}\Pi_{(t)}
        + 2d_{jkl}\pi_l
        \label{anticomsu3pi}
\end{equation}
of the Lie algebra $\mathfrak{su}(3)$, where $f_{jkl}$ and $d_{jkl}$ are the symmetric and
antisymmetric structure constants characterizing the algebra. The operators $\pi_j$ are
self-adjoint and have the traces
\begin{equation}
     {\rm Tr}\,\pi_j = 0,\quad {\rm Tr}\,\pi_j\pi_k = 32\delta_{jk}.
\end{equation}
 
The anticommutation relation (\ref{anticomsu3pi}) shows that the projection operator
$\Pi_{(t)}$ is generated from the operators $\pi_j$. The identities
\begin{equation}
    \Pi_{(a)}\,\pi_j = \delta_{at}\,\pi_j
    \label{Pipi}
\end{equation}
hold for $a=t, s$ and $j=1,\,2,\,\cdots,\,8$. Note that $\pi_j (j=1,\,2,\,\cdots,\,8)$ are
simultaneous eigen-operators of $\Pi_{(t)}$ and $\Pi_{(s)}$ with respective eigenvalues 1 and 0.
The projection operator $\Pi_{(s)}$ is orthogonal to all elements of the trio algebra.
Therefore, it is possible to afford a role to discriminate the trio and solo sectors
to the operator
\begin{equation}
     Q_X = \Pi_{(s)}
    \label{QX}
\end{equation}
which has the eigenvalue 0 and 1.

Now we are able to define the subalgebras, {\it trio algebra} $A_{(t)}$ and
{\it solo algebra} $A_{(s)}$, which can specify the trio and solo degrees of freedom by
\begin{equation}
  A_{(t)} = \{\ \Pi_{(t)}\, ; \ \pi_1, \pi_2, \cdots, \pi_8\ \}
\end{equation}
and
\begin{equation}
  A_{(s)} = \{ Q_X \}.
\end{equation}
With these algebras and the algebra of the chiral projection operator $\{L,\,R\}$, 
an algebra for the repetitional family structure, {\it family algebra} is constructed as follows: 
\begin{equation}
   A_f = \{L,\,R\}\{A_{(t)}\cup A_{(s)}\} \subset A_{in},\quad A_{(t)}\cap A_{(s)}= \emptyset.
\end{equation}
It is laborious but unmistakable to prove that the identities
\begin{equation}
        \pi_j\,A_{(a)} = \delta_{at}A_{(t)}
        \label{piA}
\end{equation}
hold for $a=t,\, s$ and $j=1, 2, \cdots, 8$. Namely, $A_{(t)}$ and $A_{(s)}$ are simultaneous
eigen-algebras of the operators $\pi_j$ with respective eigenvalues 1 and 0. We can also
verify that both algebras are irreducible under the action of the permutation group S$_3$.
 
\section{Color symmetry and B-L charge}
In the previous section, the algebra for family structure $A_f$ with the trio and solo
substructures is extracted from the internal algebra $A_{in}$. Next, it is necessary to
investigate the structure of the centralizer $C(\, A_f : A_{in}\,)$ of $A_f$ in $A_{in}$.

Let us select out the following three elements from the Dirac algebra $A_{\gm}$ as
\begin{equation}
  \rho_1 = i\gm_2\gm_3,\ \rho_2 = i\gm_3\gm_1,
  \ \rho_3 = i\gm_1\gm_2
  \label{rho}
\end{equation}
which are commutative with $\sigma_a\, (a=1, 2, 3)$ in (\ref{sigma}) and satisfy
the multiplication rules of the Pauli algebra, {\it i.e.},
$\rho_a\rho_b = \delta_{ab}I + i\ge_{abc}\rho_c$. In terms of the TDPs
of these elements, we are able to construct explicitly the centralizer $C(\, A_f : A_{in}\,)$.

The basic algebras $\{\sigma_a\}$ in (\ref{sigma}) and $\{\rho_a\}$ in (\ref{rho}) are
isomorphic with each other. Therefore, sets of their TDPs can share analogous
structures. It is, however, necessary to taken into account the point that,
while $\gm_5\otimes\gm_5\otimes\gm_5=-\Gm_5$ belongs to $A_{in}$,
the operator $\rho_3\otimes\rho_3\otimes\rho_3=-\Sigma_{12}$ is not included in $A_{in}$.
This means that no counter parts of the chiral operators can exist in the centralizer
$C(\, A_f : A_{in}\,)$. We must construct the TDPs of $\{\rho_a\}$ in
analogy with the non-chiral operators $\Pi_i$, $\Pi_{(a)}$ and $\pi_i$
to form $C(\, A_f : A_{in}\,)$. 

In accord with the operators $\Pi_i$ and $\Pi_{(a)}$, the centralizer can possess maximally
four operators which have the {\lq}{\lq}trio plus solo{\rq}{\rq} conformation. Here we
postulate that those operators carry the roles to describe the color degrees of freedom for
the tricolor quark and colorless lepton states which are established firmly in flavor physics.

In parallel with (\ref{triof}) and (\ref{solof}), it is possible to introduce the following
projection operators for the trio quark states and the solo lepton state as
\begin{equation}
   \Lam^{(q)} = \frac{1}{4}\left(3I - 1\otimes\rho_3\otimes\rho_3
   - \rho_3\otimes 1\otimes\rho_3 - \rho_3\otimes\rho_3\otimes 1\right)
    \label{trioc}
\end{equation}
and
\begin{equation}
\Lam^{(\ell)} = \frac{1}{4}\left(I + 1\otimes\rho_3\otimes\rho_3
   + \rho_3\otimes 1\otimes\rho_3 + 1\rho_3\otimes\rho_3\otimes 1\right)
   \equiv \Lam_{\ell}
    \label{soloc}
\end{equation}
which are subject to the relations
\begin{equation}
 \Lam^{(a)}\Lam^{(b)} = \delta_{ab}\Lam^{(a)}
\end{equation}
where $a, b = q, \ell$. Accordingly, we can construct the charge operator for
{\it baryon number minus lepton number} $(B-L)$ in the form
\begin{equation}
   Q_{B-L} = \frac{1}{3}\Lam^{(q)} - \Lam^{(\ell)}
        = - \frac{1}{3}(1\otimes\rho_3\otimes\rho_3
          + \rho_3\otimes 1\otimes\rho_3 
          + \rho_3\otimes\rho_3\otimes 1)
    \label{B-L}
\end{equation}
which has eigenvalues 1/3 and -1. It is possible to subdivide the operator $\Lam^{(q)}$
symmetrically into
\begin{equation}
      \Lam^{(q)} = \Lam_r + \Lam_y + \Lam_g
\end{equation}
where
\begin{equation}
 \left\{\ 
  \begin{array}{l}
   \Lam_r = \frac{1}{4}\left(I + 1\otimes\rho_3\otimes\rho_3
   - \rho_3\otimes 1\otimes\rho_3 - \rho_3\otimes\rho_3\otimes 1
   \right), \\
   \noalign{\vskip 0.3cm}
   \Lam_y = \frac{1}{4}\left(I - 1\otimes\rho_3\otimes\rho_3
   + \rho_3\otimes 1\otimes\rho_3 - \rho_3\otimes\rho_3\otimes 1
   \right), \\
   \noalign{\vskip 0.3cm}
   \Lam_g = \frac{1}{4}\left(I - 1\otimes\rho_3\otimes\rho_3
   - \rho_3\otimes 1\otimes\rho_3 + \rho_3\otimes\rho_3\otimes 1
   \right)\\
     \noalign{\vskip 0.1cm}
  \end{array}
  \right.
\end{equation}
can be interpreted as the projection operators into tricolor quark modes.
The four operators $\Lam_a$ satisfy the relations
\begin{equation}
    \Lam_a\Lam_b = \delta_{ab}\Lam_a,\ \ \sum_{a}\Lam_{a}=I
\end{equation}
for $a,\,b = r,\,y,\,g,\,\ell$.

Here again, taking resort to the isomorphism of the basic algebras $\{\sigma_a\}$ in
(\ref{sigma}) and $\{\rho_a\}$ in (\ref{rho}), we construct the generators of
transformations among the tricolor quark states in analogy with (\ref{pi}) as follows:
\begin{equation}
\left\{\ 
\begin{array}{l}
    \lam_1 = \half\left(\rho_1\otimes\rho_1\otimes 1
          + \rho_2\otimes\rho_2\otimes 1\right),\\
          \noalign{\vskip 0.3cm}
    \lam_2 = \half\left(\rho_1\otimes\rho_2\otimes\rho_3
          - \rho_2\otimes\rho_1\otimes\rho_3\right),\\
          \noalign{\vskip 0.3cm}
    \lam_3 = \half\left(1\otimes\rho_3\otimes\rho_3
          - \rho_3\otimes 1\otimes\rho_3\right),\\
          \noalign{\vskip 0.3cm}
    \lam_4 = \half\left(\rho_1\otimes 1\otimes\rho_1
          + \rho_2\otimes 1\otimes \rho_2\right), \\
          \noalign{\vskip 0.3cm}
    \lam_5 = \half\left(\rho_1\otimes\rho_3\otimes\rho_2
          - \rho_2\otimes\rho_3\otimes\rho_1\right),\\
          \noalign{\vskip 0.3cm}
    \lam_6 = \half\left(1\otimes\rho_1\otimes\rho_1
          + 1\otimes\rho_2\otimes\rho_2\right),\\
            \noalign{\vskip 0.3cm} 
    \lam_7 = \half\left(\rho_3\otimes\rho_1\otimes\rho_2
          - \rho_3\otimes\rho_2\otimes\rho_1\right),\\
          \noalign{\vskip 0.3cm}
    \lam_8 = \frac{1}{2\sqrt{3}}\left(1\otimes\rho_3\otimes\rho_3
          + \rho_3\otimes 1\otimes\rho_3 
          -2\rho_3\otimes\rho_3\otimes 1\right)    
\end{array}
\right.
\label{lam}
\end{equation}
which satisfy the commutation relations
\begin{equation}
        [\lam_j,\ \lam_k] = 2f_{jkl}\lam_l
        \label{comsu3lam}
\end{equation}
and the anticommutation relations
\begin{equation}
       \{\lam_j,\ \lam_k\} = \frac{4}{3}\delta_{jk}\Lam^{(q)}
        + 2d_{jkl}\lam_l 
        \label{anticomsu3lam}
\end{equation}
of the Lie algebra $\mathfrak{su}(3)$.

The operators $\lam_j$ are self-adjoint and have the traces
\begin{equation}
     {\rm Tr}\lam_j = 0,\quad {\rm Tr}\lam_j\lam_k = 32\delta_{jk}.
\end{equation}
Corresponding to (\ref{Pipi}), the operators $\Pi^{(a)}$ and $\pi_j$ satisfy the identities
\begin{equation}
        \Lam^{(a)}\lam_j = \delta_{aq}\lam_j
\end{equation}
which hold for $a=q, \ell$ and $j=1, 2, \cdots, 8$. Evidently, the operators $\lam_j$
works to annihilate the leptonic mode and the operator $\Lam^{(\ell)}$ is orthogonal to
all elements $\lam_j$ of the trio quark algebra.
Now we are able to define the subalgebras, {\it trio quark algebra} $A^{(q)}$ and
{\it solo lepton algebra} $A^{(\ell)}$, which can specify the trio and solo degrees of freedom by
\begin{equation}
  A^{(q)} = \{\ \Lam^{(q)}\,\ ; \ \lam_1, \lam_2, \cdots, \lam_8\ \}
\end{equation}
and
\begin{equation}
  A^{(\ell)} = \{ \Lam^{(\ell)}=\Lam_{\ell} \}.
\end{equation}
With these algebras, an algebra for the four color degrees of freedom, {\it color algebra} is
constructed as follows: 
\begin{equation}
   A_c = A^{(q)}\cup A^{(\ell)} \subset A_{in},\quad A^{(q)}\cap A^{(\ell)}= \emptyset.
\end{equation}
Just as (\ref{piA}), the operators $\lam_j$ and the algebras $A^{(q)}$ and $A^{(\ell)}$ satisfy
the identities
\begin{equation}
        \lam_jA^{(a)} = \delta_{aq}A^{(q)}
        \label{lambdaA}
\end{equation}
hold for $a=q,\, \ell$ and $j=1, 2, \cdots, 8$. Namely, $A^{(q)}$ and $A^{(\ell)}$ are
simultaneous eigen-algebras of the operators $\lam_j$. Both algebras are proved to be
irreducible under the action of the permutation group S$_3$.

\section{Triplet field: Multiple spinor representation of the Lorentz group}
With these formation of the external and internal algebras, we are now able to introduce
the triplet field $\Psi(x)$, on the spacetime point $x^\mu$, which behave like the
TDP of four dimensional Dirac spinor. The triplet field which span
$4^3$ dimensional vector space possesses $4^2$ component fields with spin $\half$ with
internal attributes of color and family degrees of freedom. 

With the triplet field $\Psi(x)$ and its adjoint field
\begin{equation}
  \overline{\Psi}(x) = \Psi^\dagger(x)\Gm_0,
\end{equation}
the Lorentz invariant scalar product is defined as
\begin{equation}
  \overline{\Psi}(x)\Psi(x)
  = \sum_{abc}\overline{\Psi}_{abc}(x)\Psi_{abc}(x).
  \label{Linvariant}
\end{equation}
Under the proper Lorentz transformation
$x^{\prime \mu} = \Omega^\mu{}_\nu x^\nu$ where
$\Omega_{\lambda\mu}\Omega^\lambda{}_\nu = \eta_{\mu\nu}$
and ${\rm det}\,\Omega = 1$, the triplet field and its adjoint are transformed as
\begin{equation}
  \Psi^\prime(x^\prime) = S(\Omega)\Psi(x),\ \ 
  \overline{\Psi}^\prime(x^\prime) = \overline{\Psi}(x)S^{-1}(\Omega)
\end{equation}
where the transformation matrix is explicitly given by
\begin{equation}
   S(\Omega) =
   \exp\left(-\frac{i}{2}M_{\mu\nu}\omega^{\mu\nu}\right)
\end{equation}
with the generator for the spacetime rotations in (\ref{Lorentzgenerator}) and
the angles $\omega^{\mu\nu}$ in the $\mu$-$\nu$ planes.
For the discrete spacetime transformations such as the space inversion, the time reversal
and the charge conjugation, the present scheme retains exactly the same structure as
the ordinary Dirac theory.

At the level of the algebra, the family and color degrees of freedom have dual structures:
the part $A_{(t)}\cup A_{(s)}$ of the family algebra is isomorphic to the color algebra
$A_c = A^{(q)}\cup A^{(\ell)}$. This duality is not realized in the observed characteristics
of fundamental fermions. The breakdown of the duality existing in the algebras arises at the
level of continuous group formations. Here we introduce the group of color symmetry by
\begin{equation}
     SU_c(3) = \left\{ \exp\left(-\frac{i}{2}\sum_j\lam_j\theta^{j}(x)\right) \right\}
\end{equation}  
and the group of $Q_{B-L}$ charge and $Q_X$ charge, respectively, by
\begin{equation}
  U_{B-L}(1) = \left\{ \exp\left(-\frac{i}{2}Q_{B-L}\theta_{B-L}(x)\right) \right\},\quad
  U_{X}(1) = \left\{ \exp\left(-\frac{i}{2}Q_{X}\theta_{X}(x)\right) \right\}
\end{equation}
where $\theta^{j}(x)$, $\theta_{B-L}(x)$ and $\theta_{X}(x)$ are arbitrary real functions
of space-time. Under the action of these groups, the bilinear Lorentz invariants in
(\ref{Linvariant}) is postulated to be invariant.

The trio plus solo construction of the internal algebra renders the triplet field $\Psi(x)$
to possess the following decompositions as
\begin{equation}
\begin{array}{lll}
  \Psi(x)&=&\Psi^{(q)}(x) +  \Psi^{(\ell)}(x)\\
  \noalign{\vskip 0.3cm}
   &=&\Psi_{(t)}(x) +  \Psi_{(s)}(x)\\
  \noalign{\vskip 0.3cm}
   &=&\Psi^{(q)}_{(t)}(x) + \Psi^{(q)}_{(s)}(x) +  \Psi^{(\ell)}_{(t)}(x) + \Psi^{(\ell)}_{(s)}(x)\\
\end{array}
\end{equation}
with
\begin{equation}
  \Psi^{(a)}_{(b)}(x) = \Lam^{(a)}\Psi_{(b)}(x) = \Pi_{(b)}\Psi^{(a)}(x)
 = \Lam^{(a)}\Pi_{(b)}\Psi(x)
\end{equation}
where $a=q,\,\ell$ and $b=t,\,s$. Accordingly, the bilinear Lorentz invariant in (\ref{Linvariant})
is expressed in terms of the separate invariants of the trio and solo sectors as
\begin{equation}
  \overline{\Psi}(x)\Psi(x)
  = \sum_{a=l,\ell}\overline{\Psi}^{(a)}(x)\Psi^{(a)}(x)
  = \sum_{b=t,s}\overline{\Psi}_{(b)}(x)\Psi_{(b)}(x)
  = \sum_{a=l,\ell;b=t,s}\overline{\Psi}^{(a)}_{(b)}(x)\Psi^{(a)}_{(b)}(x).
\end{equation}

To extract the component fields of internal modes of the triplet field, it is convenient
to introduce the braket symbols for the projection operators $\Lam_a$ and $\Pi_{ih}$ by
\begin{equation}
    \Lam_a = \mid a\rangle\langle a\mid,\ \ 
    \Pi_{ih} = \mid i\,h\rr\ll i\,h\mid .
\end{equation}
For the triplet field and its conjugate, the projection operators work as follows:
\begin{equation}
  \Lam_a\Pi_{ih}\Psi(x) = \Lam_a|ih\rr\ll ih|\Psi(x)\rangle
                        = |aih\rr\ll aih|\Psi(x)\rangle
                        = |aih\rr\Psi_{aih}(x)
\end{equation}
and
\begin{equation} 
 \bar{\Psi}(x)\Lam_a\Pi_{ih} = \langle\bar{\Psi}(x)\Lam_a|ih\rr\ll ih| = \langle\bar{\Psi}(x)|aih\rr\ll aih|
 = \bar{\Psi}_{ai\bar{h}}(x)\ll aih|
\end{equation}
where $\Psi_{aih}(x)$ and $\bar{\Psi}_{aih}(x)$ are chiral component fields, and $\bar{h}$
implies that $\bar{L}=R$ and $\bar{R}=L$. Then, the decomposition of the bilinear scalar and vector forms of the triplet fields can be achieved as follows:
\begin{equation}
  \bar{\Psi}(x)\Psi(x)
  =\sum_{a}\sum_{ih}\bar{\Psi}(x)\Lam_a\Pi_{ih}\Psi(x)
  =\sum_{a}\sum_{ih}\bar{\Psi}_{ai\bar{h}}(x)\Psi_{aih}(x)
  \label{scalarbillinear}
\end{equation}
and
\begin{equation}
  \bar{\Psi}(x)\Gm_\mu\Psi(x)
  = \sum_{a}\sum_{ih}\bar{\Psi}(x)\Gm_\mu\Lam_a\Pi_{ih}\Psi(x)
  = \sum_{a}\sum_{ih}\bar{\Psi}_{aih}(x)\Gm_\mu\Psi_{aih}(x).
  \label{vecterbillinear}
\end{equation}

\section{Standard model in the triplet field formalism\label{standard}}
To incorporate the Weinberg-Salam mechanism in the present scheme, it is necessary
to prepare a set of triplet fields which form representations of the symmetry group
SU$_L(2)\times$U$_y(1)$.

We introduce the left-handed triplet field constituting the doublet of electroweak isospin
$\half \tau_a$ with $y=0$ as
\begin{equation}
  \Psi_L(x) =
     \left(  
       \begin{array}{ccc}
          \Psi_u(x)\\
          \noalign{\vskip 0.2cm}
          \Psi_d(x)\\
       \end{array}
     \right)_L
  \label{chiraldoublet}
\end{equation}
where $\Psi_u(x)$ and $\Psi_d(x)$ signify the up and down isospin components, respectively,
and the right-handed triplet fields forming the electroweak isospin singlets with $y=1$ and
$y=-1$ as follows:
\begin{equation}
          \Psi_U(x),\quad \Psi_D(x).
  \label{chiralsinglets}         
\end{equation}
The chiral fabric of these triplets claims the conditions: $R\Psi_u=R\Psi_d=0$ and
$L\Psi_U=L\Psi_D=0$. The breakdown of the electroweak symmetry is assumed to be triggered
by the ordinary Higgs field:
\begin{equation}
     \phi(x) =
      \left(
       \begin{array}{c}
        \phi^+(x) \\
        \noalign{\vskip 0.2cm}
        \phi^0(x) \\
       \end{array}
      \right),
      \quad
      \tilde{\phi}(x) =
      \left(
       \begin{array}{c}
        \phi^{0\ast}(x) \\
        \noalign{\vskip 0.2cm}
        -\phi^-(x) \\
       \end{array}
      \right) . 
      \label{ordinaryHiggs}    
\end{equation}

The quark and lepton parts of the electroweak doublet $\Psi_L(x)$ have the component modes
expressed schematically as
\begin{equation}
  \Psi^{(q)}(x) = \Lam^{(q)}\Psi_L(x) =
     \left(  
       \begin{array}{ccc}
          \Lam^{(q)}\Lam_a\Pi_{iL}\Psi_u\\
          \noalign{\vskip 0.2cm}
          \Lam^{(q)}\Lam_a\Pi_{iL}\Psi_d\\
       \end{array}
     \right)_L =
     \left(  
       \begin{array}{cccc}
        u_a\ & c_a\ & t_a & \cdot\\
        \noalign{\vskip 0.2cm}
        d_a\ & s_a\ & b_a & \cdot
       \end{array}
     \right)_L
\end{equation}
and
\begin{equation}
  \Psi^{(\ell)}(x) = \Lam^{(\ell)}\Psi_L(x) =
    \left(  
     \begin{array}{ccc}
       \Lam^{(\ell)}\Pi_{iL}\Psi_u\\
       \noalign{\vskip 0.2cm}
       \Lam^{(\ell)}\Pi_{iL}\Psi_d\\
     \end{array}
    \right)_L
  = \left(  
     \begin{array}{cccc}
      \nu_e\ & \nu_\mu\ & \nu_\mu & \cdot\\
      \noalign{\vskip 0.2cm}
      e\ & \mu\ & \tau & \cdot
     \end{array}
    \right)_L 
\end{equation}
where the symbol $\,\cdot\,$ expresses the fourth fermions with $Q_X=1$ charge which remain
anonymous at present. Similarly, the electroweak singlets $\Psi_U(x)$ and $\Psi_D(x)$ possess
the quark and lepton parts with contents of the component modes as follows:
\begin{equation}
\ \Psi^{(u)}(x) = \Lam^{(q)}\Psi_U(x)
 = \left(\Lam^{(q)}\Lam_a\Pi_{iR}\Psi_U\right)
 = \left(  
    \begin{array}{cccc}
     u_a\ & c_a\ & t_a & \cdot\\
    \end{array}
   \right)_R ,
\end{equation}
\begin{equation}
\ \Psi^{(d)}(x) = \Lam^{(q)}\Psi_D(x)
 = \left(\Lam^{(q)}\Lam_a\Pi_{iR}\Psi_D\right)
 = \left(  
    \begin{array}{cccc}
     d_a\ & s_a\ & b_a & \cdot\\
   \end{array}
   \right)_R ,
\end{equation}
\begin{equation}
  \Psi^{(\nu)}(x) = \Lam^{(\ell)}\Psi_U(x)
 = \left(\Lam^{(\ell)}\Pi_{iR}\Psi_U\right)
 = \left(  
    \begin{array}{cccc}
     \nu_e\ & \nu_\mu\ & \nu_\tau & \cdot\\
    \end{array}
   \right)_R ,\ \ 
\end{equation}
and
\begin{equation}
\Psi^{(e)}(x) = \Lam^{(\ell)}\Psi_D(x)
 = \left(\Lam^{(\ell)}\Pi_{iR}\Psi_D\right)
 = \left(  
    \begin{array}{cccc}
     \ e\,\ \ & \mu\,\ \ & \tau & \cdot\\
    \end{array}
   \right)_R .\ \ 
\end{equation}

In terms of the chiral triplet fields in (\ref{chiraldoublet}) and (\ref{chiralsinglets}),
the Lagrangian density of the kinetic part including the gauge interactions of the fundamental
fermions is given in the generic form
\begin{equation}
  {\cal L}_{kg} = \bar{\Psi}_L(x)\Gm^\mu{\cal D}_\mu\Psi_L(x)
  + \bar{\Psi}_U(x)\Gm^\mu{\cal D}_\mu\Psi_U(x)
  + \bar{\Psi}_D(x)\Gm^\mu{\cal D}_\mu\Psi_D(x)
  \label{kineticgauge}
\end{equation}
with the covariant derivatives ${\cal D}_\mu$ which act on the triplet fields as follows:
\begin{equation}
{\cal D}_\mu\Psi_L(x)=
\left(\pd_\mu - gA_\mu^{a}(x)\half\tau_a
              - g^{\pr}B_\mu(x)\half Y
              - g_cA_{c\mu}^{a}(x)\half\lam_a 
              - g_X\,X_{c\mu}^{a}(x)\half Q_X \right)\Psi_L(x),
\end{equation}
\begin{equation}
{\cal D}_\mu\Psi_U(x)=
\left(\pd_\mu - g^{\pr}B_\mu(x)\half Y
       - g_cA_{c\mu}^{a}(x)\half\lam_a 
       - g_X\,X_{c\mu}^{a}(x)\half Q_X \right)\Psi_U(x)
\end{equation}
and
\begin{equation}
{\cal D}_\mu\Psi_D(x)=
\left(\pd_\mu - g^{\pr}B_\mu(x)\half Y
      - g_c\,A_{c\mu}^{a}(x)\half\lam_a 
      - g_X\,X_{c\mu}^{a}(x)\half Q_X \right)\Psi_D(x)
\end{equation}
where $A_\mu^{a}(x)$, $B_\mu(x)$, $A_{c\mu}^{a}(x)$ and $X_\mu(x)$ are respectively the gauge
fields for the electroweak isospin, hypercharge $Y=y+Q_{B-L}$, color symmetry and U$_X$(1)
symmetry. This density ${\cal L}_{kg}$ is readily proved to reproduce the gauge interactions
of the standard model which includes also the right-handed neutrino species and the fourth
additional family of fermions with $X$ charge. 

In the present scheme, the chiral triplet fields include the repetitional degrees of freedom for family structure. This feature enables us to formulate the Yukawa interactions in an intrinsic mechanism without bringing in so many unknown parameters from outside. In terms of chiral triplet fields and the Higgs field, the Lagrangian density of the Yukawa interactions can be expressed
to be
\begin{equation}
  \begin{array}{ll}
   {\cal L}_Y &= g_u\bar{\Psi}^{(q)}(x){\cal Y}_u
            \tilde{\phi}(x) \Psi^{(u)}(x)
            + g_d\bar{\Psi}^{(q)}(x){\cal Y}_d
            \phi(x)\Psi^{(d)}(x)\\
             \noalign{\vskip 0.4cm}
             &+\ g_\nu\bar{\Psi}^{(\ell)}(x){\cal Y}_\nu
            \tilde{\phi}(x)\Psi^{(\nu)}(x)
              + g_e\bar{\Psi}^{(\ell)}(x){\cal Y}_e
            \phi(x)\Psi^{(e)}(x) + {\rm h.c.}
  \end{array}
\label{Yukawaint}
\end{equation}
where the kernel operator ${\cal Y}_a\ (a=u,\,d,\,\nu,\,e)$ works to induce mixings among
repetitional family modes. To be commutative with the generators of color symmetry, the operator
must be composed solely of the elements of the family algebra $A_f$. The quark part of the
density ${\cal L}_Y$ is decomposed into the Yukawa interactions among component quark fields as
\begin{equation}
  g_a\bar{\Psi}^{(q)}{\cal Y}_a\Psi^{(a)} =
  g_a\sum_{ij}\bar{\Psi}^{(q)}\Pi_{iR}{\cal Y}_a\Pi_{jR}\Psi^{(a)}
  = \sum_{ij}\bar{\Psi}_{i}^{(q)}
  \left[g_a\ll iR\mid{\cal Y}_a\mid jR\rr \right]\Psi_{j}^{(a)}
\end{equation}
where $a=u,\,d$. This result and the similar decomposition of the lepton part lead us to
the interpretation that the quantities
\begin{equation}
      g_a\ll i|{\cal Y}_a|j\rr \equiv g_a\ll iR|{\cal Y}_a|jR\rr
      \label{Yukawacoupling}
\end{equation}
are the Yukawa coupling constants of the $a$ sector $(a=u,\,d,\,\nu,\,e)$.

Let us express the kernel operators ${\cal Y}_a$ generically by ${\cal Y}$. In general,
${\cal Y}$ can consist of arbitrary elements of the trio and solo algebras $A_{(t)}$ and
$A_{(s)}$. At present we have no principle to restrict its form. It is experimental
information in flavor physics which work to restrict the composition of the kernel operator.
Namely, we must repeat trial and error of data fittings with candidates of kernel operator
whose compositions are fixed under possible working hypotheses. Here, we restrict our
consideration to the ordianry trio families and impose one simple hypothesis that
the kernel is self-adjoint and composed of the following components as
\begin{equation}
  {\cal Y}\ =\ {\cal Y}_{(t)\,sym} + {\cal Y}_{(t)\,asym}
    \label{kernel}
\end{equation}
where ${\cal Y}_{(t)\,sym}$ and ${\cal Y}_{(t)\,asym}$ are, respectively, composed of such
elements of $A_{(t)}$ that are symmetric and asymmetric parts under the cyclic permutation
of the S$_3$. The symmetric and asymmetric parts are determined to be
\begin{equation}
\begin{array}{lll}
{\cal Y}_{(t)\,sym} & = & 2a\,(\pi_2-\pi_5+\pi_7) + 2b\,(\pi_1+\pi_4+\pi_6)
             + c\,\Pi_{(t)} \\
  \noalign{\vskip 0.4cm}
                    & = & a\,[\ \sg_1\otimes\sg_2\otimes\sg_3
                  + \sg_3\otimes\sg_1\otimes\sg_2
                  + \sg_2\otimes\sg_1\otimes\sg_3 \\
  \noalign{\vskip 0.3cm}
            & &\ \  - \sg_3\otimes\sg_2\otimes\sg_1
                   - \sg_1\otimes\sg_3\otimes\sg_2
                   - \sg_2\otimes\sg_1\otimes\sg_3\ ] \\
 \noalign{\vskip 0.4cm}
           & + & b\,[\ 1\otimes\sg_1\otimes\sg_1
                  + \sg_1\otimes 1\otimes\sg_1
                  + \sg_1\otimes\sg_1\otimes 1 \\
 \noalign{\vskip 0.3cm}
            & &\ \  +  1\otimes\sg_2\otimes\sg_2
                   + \sg_2\otimes 1\otimes\sg_2
                   + \sg_2\otimes\sg_2\otimes 1\ ] + c\,\Pi_{(t)}
\end{array}
\label{symmetric}
\end{equation}
and
\begin{equation}
 \begin{array}{lll}
{\cal Y}_{(t)\,asym} & = & e\,(\pi_1-\pi_6) \\
  \noalign{\vskip 0.4cm}
           & = & e\,[\sg_1\otimes\sg_1\otimes 1+ \sg_2\otimes\sg_2\otimes 1
                          - 1\otimes\sg_1\otimes\sg_1 - 1\otimes\sg_2\otimes\sg_2].
\end{array}
    \label{asymmetric}
\end{equation}
All coefficients in these operators are real constants. The forms of the operators
${\cal Y}_{(t)\,sym}$ in (\ref{symmetric}) are unique.
We choose the operator ${\cal Y}_{(t)\,asym}$ in (\ref{asymmetric}) from many other
possibility so as to simplify numerical analyses of Dirac mass matrices. 

To calculate the Yukawa coupling constants as the matrix elements in (\ref{Yukawacoupling}),
let us specify matrix representations of the chiral projection operators in $A_\gm$ as
\begin{equation}
\begin{array}{ll}
   \ell = \frac{1}{2}(1-\gm_5)
        = \frac{1}{2}
           \left(
           \begin{array}{rr}
              1  &   -1 \\
             -1  &    1
           \end{array}
           \right), & 
   r = \frac{1}{2}(1+\gm_5)
       = \frac{1}{2}
           \left(
           \begin{array}{cc}
              1  &    1 \\
              1  &    1
           \end{array}
           \right) .
\end{array}
\end{equation}
The eigenvectors of these operators are given in the forms
\begin{equation}
   \begin{array}{ll}
        \mid\! \ell\, \rr = \frac{1}{\sqrt{2}}
                       \left(
                         \begin{array}{c}
                              \mid \psi \rangle \\
                          \noalign{\vskip 0.2cm}
                            -\mid \psi\rangle
                         \end{array}
                        \right)
                        e^{i\gth},\quad
           \mid\! r\, \rr = \frac{1}{\sqrt{2}}
                       \left(
                         \begin{array}{c}
                              \mid \psi \rangle \\
                          \noalign{\vskip 0.2cm}
                              \mid \psi\rangle
                         \end{array}
                        \right)
      \end{array}
      \label{eigenvector}
\end{equation}
where $\mid\psi\,\rangle$ is a normalized Pauli spinor and $\gth$ is an arbitrary phase.
In terms of the ordered products of these component eigenvector, we are able to construct
the state vectors of interaction modes $\mid j\,\rr=\mid jR\,\rr$ as follows:
\begin{equation}
    \mid 1\,\rr=\,\mid\! r\ell\ell \rr,\ \ \mid 2\,\rr=\,\mid\! \ell r\ell \rr,\ \ 
    \mid 3\,\rr=\,\mid\! \ell\ell r\rr,\ \ \mid 4\,\rr=\,\mid\! rrr \rr
\end{equation}
with $\mid\! r\ell\ell \rr = \mid\! r \rr_1\mid\! \ell \rr_2\mid\ell \rr_3$ etc. Then,
careful calculation of the matrix elements of the kernel operator in (\ref{kernel})
results in
\begin{equation}
\begin{array}{lll}
(\, \ll i\mid\! {\cal Y}\!\mid j\rr\, ) &\propto&
\quad\left(
\begin{array}{c|l}
  & |r\ell\ell\rr \ |\ell r\ell\rr \ |\ell\ell r\rr \ | rrr\rr \\
 \hline
  \ll{r\ell\ell}|  &     \\
  \ll{\ell r\ell}| & \qquad\quad {\rm Kernel} \quad  \\
 \ll{\ell\ell r}|  & \qquad\qquad {\cal Y}\qquad   \\
  \ll{rrr}   |    &     \\
 \end{array}
\right) \\
 \noalign{\vskip 0.5cm}
 &=&
\left(
 \begin{array}{cccc}
          c            & (ia+b+e)\, e^{i\ga_3}  & (-ia + b)\, e^{i\ga_2}  & 0  \\
\noalign{\vskip 0.3cm}
  (-ia+b+e)\, e^{-i\ga_3} &         c            & (ia+b-e)\, e^{i\ga_1}  & 0 \\
 \noalign{\vskip 0.3cm}
  (ia+b)\, e^{-i\ga_2} & (-ia+b-e)\, e^{-i\ga_1} &           c          & 0 \\
\noalign{\vskip 0.3cm}
          0            &         0               &           0          & 0 \\
 \end{array}\ 
\right)
\end{array}
\label{YukawacouplingMatrix}
\end{equation}
where $\ga_{j}=\gth_{j+1}-\gth_{j+2}\ (\gth_{j+3}=\gth_j)$ is the difference
of the phases in the component ket vectors $\mid\! \ell \rr_{j+1}$ and
$\mid\ell \rr_{j+2}$\ \,$(\mid\ell \rr_{j+3}=\mid\ell \rr_{j})$.
Due to the orthogonality between the trio and solo algebras, there exist no Yukawa coupling
constant which works to mix the ordinary trio and additional solo families of fundamental
fermions.

In the present theory of fundamental fermions, we inherit the Weinberg-Salam mechanism for
the breakdown of the electroweak symmetry. At and below the energy scale where
the Higgs field takes the vacuum expectation value, the Yukawa interaction (\ref{Yukawaint})
results in the Dirac mass matrices. For the charged sector of the trio families, we obtain,
generically, the following Hermitian matrix.
\begin{equation}
    {\cal M}  \propto
 P^\dagger
   \, \left(
      \begin{array}{ccc}
        c          &   i a  + b + e & - i a + b    \\
      \noalign{\vskip 0.4cm}
     - i a + b + e &         c      &  i a +b -e  \\
      \noalign{\vskip 0.4cm} 
      i a + b &  - i a + b - e &      c       \\
     \end{array}
\right)\,P
\label{DiracMassMatrix}
\end{equation}
where
\begin{equation}
 P= {\rm diagonal}\left( e^{-i\gth_1},\ e^{-i\gth_2},\ e^{-i\gth_3} \right)
\end{equation}
is the diagonal phase matrix. The charged solo fermion in the isolated fourth mode has
the mass proportional to the constant $f$.

\section{Discussion}
In this way, we have formulated an algebraic theory for description of external and internal
attributes and symmetries of fundamental fermions. The basic ingredients of the theory are
the {\it triplet algebra} consisting of all the TDPs of the Dirac matrices
and the {\it triplet field} forming the multiple spinor representation of the Lorentz group.

The triplet algebra is decomposed into the product of the external and internal algebras. All
elements of the external algebra $A_{ex}$ are invariant under the action of permutation group
S$_3$ which works to exchange the order of the $A_\gm$ elements in the TDP.
The internal algebra is the product of the subalgebras of the family and color degrees
of freedom, $A_f$ and $A_c$. Both subalgebras have the trio plus solo conformation as
$A_f = \{L,\,R\}\{A_{(t)}\cup A_{(s)}\}$ and $A_c = A^{(q)}\cup A^{(\ell)}$. All of the trio
and solo subalgebras, $A_{(t)}$, $A^{(q)}$, $A_{(s)}$ and $A^{(\ell)}$, are irreducible under
the group S$_3$. On these algebraic constructions, the triplet field possesses the internal
group structure of the color symmetry and the $B-L$ and $X$ charges.

The Weinberg-Salam theory of electroweak interaction is formulated in the triplet field
formalism by postulating the existence of a set of the triplet fields of the electroweak
isospin doublet and singlets as in (\ref{chiraldoublet}) and (\ref{chiralsinglets}). It
should be noticed that the $y$ charge has only a temporal role to distinguish the singlet
fields $\Psi_U$ and $\Psi_D$ and do not have its own gauge field. In the Lagrangian density
in (\ref{kineticgauge}), the ordinary gauge field $B_{\mu}(x)$ of U$_Y$(1) symmetry is
assumed to couple to the hypercharge $Y=y+Q_{B-L}$. It is tempting to develop a left and
right symmetric extension of electroweak interaction and to relate the $y$ charge to
a component of a right-handed isospin. However, we leave the study of such possibility in
future as an open problem. 

Most characteristic aspect of the triplet field formalism lies in the points that fundamental 
fermions are identified with component fields of the triplet fields and interactions among
the fermions can be interpreted as transitions among component fields. The gauge and Yukawa
interactions of the standard model are arranged, respectively, in the Lagrangian densities
${\cal L}_{kg}$ in (\ref{kineticgauge}) and ${\cal L}_Y$ in (\ref{Yukawaint}) with diagonal
and off-diagonal combinations of chiral triplet fields. The trio plus solo formations of
family modes enables us to separate the interactions of ordinary fermions and those of the
additional fourth fermions. Using the kernel in (\ref{kernel}), we are able to {\it calculate}
the Yukawa coupling constants as its matrix elements. Note that the phase of CP violation
originates from the phases in the eigenvectors in (\ref{eigenvector}).

The mass matrix of Dirac type in (\ref{DiracMassMatrix}) has new and unique structure.
If the parameters $b$ and $c$ take close and dominant values, it becomes the well known
mass matrix of democratic type. With this mass matrix, we can carry out numerical analyses
on the mass spectra and the weak mixing matrix for the ordinary quark sector and succeed
to explain all data within experimental accuracy. Results of numerical analyses will be
published elsewhere.

For further investigations of the lepton sector 
$\Psi^{(\ell)}=\Psi^{(\ell)}_{(t)}+\Psi^{(\ell)}_{(s)}$ and the solo family sector 
$\Psi_{(s)}=\Psi^{(q)}_{(s)}+\Psi^{(\ell)}_{(s)}$, it is necessary to enrich contents
of bosonic fields. So far, only one kind of Higgs field $\phi(x)$ of the standard model
in (\ref{ordinaryHiggs}) is chosen to describe the Weinberg-Salam mechanism.
The main reason for this restriction is to severely suppress the appearance of the
flavor-changing-neutral-current. To explain the smallness of the neutrino masses,
however, it is required to introduce a different type of scalar field which brings about
the seesaw mechanism by creating the so-called Majorana masses. Further, we must introduce
an extra Higgs field which works to break down the U$_X(1)$ symmetry and metamorphose
the gauge field $X_\mu(x)$ to a vector field with a sufficiently large mass. This Higgs
field is assumed to couple solely to the solo family so that the fourth fermions acqurie
sufficiently heavy masses compared with the ordinary trio fermions. We can expect that
physics of the solo family sector can naturally be related to rapidly developing LHC
experiments and astrophysics. There exist plausible candidates for the dark matter in
the hitherto-unobserved solo family sector. Such possibilities must be pursued in the
triplet field formalism.        

Last but not least. In this article, we have identified the internal algebra $A_{in}$ with
the centralizer of the algebra for the Lorentz transformation in (\ref{Ain}) as
\[
   A_{in} = C( \{\Sigma_{\mu\nu}: A_T \})= \langle\ 1\otimes \gm_\mu\otimes \gm_\mu,\  
   \gm_\mu\otimes 1\otimes\gm_\mu,\ \gm_5\otimes\gm_5\otimes\gm_5\ \rangle. 
\]
This choice turns out convenient to find the algebra of the family degrees of freedom
through the decompositions of the chiral projection operators in (\ref{Loperator}) and
(\ref{Roperator}) which are impossible in the external algebra $A_{ex}$. However, once
we found the projection operators $\Pi_j$ to the four family modes and the relations
in (\ref{nonchiraltochiral}) which allow to regenerate the chiral operators $\Pi_{jh}$,
it is possible to use the centralizer of the external algebra   
\begin{equation}
   A_{in} = C( \{ A_{ex}: A_T \})= \langle\ 1\otimes \gm_\mu\otimes \gm_\mu,\  
   \gm_\mu\otimes 1\otimes\gm_\mu\ \rangle 
   \label{noGm5}
\end{equation}
as an alternative definition of the internal algebra. Mathematically, this simple choice
which satisfies the relation
\begin{equation}
   A_T = A_{ex} A_{in},\quad  A_{ex}\cap A_{in} = \emptyset
      \label{noGm5}
\end{equation}
seems relevant since it is faithful to the Coleman-Mandula theorem\cite{Coleman,Weinberg}
insisting that the {\lq\lq}spacetime and internal symmetries cannot be combined in any but
a trivial way{\rq\rq}. However, both algebras have no essential difference in physics. 

\ack
This work was supported by the JSPS Grant-in-Aid for Scientific Research (S) No.\,22224003.

\end{document}